\documentclass[a4paper]{jpconf}
\usepackage{amssymb}
\usepackage{amsthm}
\usepackage{latexsym}
\usepackage{graphics}
\usepackage{psfrag,fancyhdr,epsfig}
\newcommand{\be}{\begin{eqnarray}}
\newcommand{\ee}{\end{eqnarray}}
\newcommand{\bdm}{\begin{displaymath}}
\newcommand{\edm}{\end{displaymath}}

\begin{document}
\title{\textbf{Exact Solutions in 5-D Brane Models With Scalar Fields} }
\author{\textbf{C. Bogdanos\footnote{Work done in collaboration with A. Dimitriadis
and K. Tamvakis}}}
\date{}
\address{Physics Department, University of Ioannina\\
Ioannina GR451 10, Greece}
 \vspace{1cm}
%\numberwithin{equation}{section}
%%%%%%%%%%%%%%%%%%%%%%%%%%%%%%%%%%%%%%%%%%%%%%%%%%%%%%%
\begin{abstract}
In this talk we consider the problem of a scalar field, non-minimally coupled to gravity through a $-\xi\phi^{2}R$ term, in
the presence of a Brane. A number of exact solutions, for a wide range of values of the coupling parameter $\xi$,
for both $\phi$-dependent
and $\phi$-independent Brane tension, are presented. The behavior and general features of these solutions are discussed.
We derive solutions for the scalar field compatible with the Randall-Sundrum metric and also geometries which can accomodate a folded kink-like scalar.
Analytic and numerical results are provided
for the case of a Brane or for smooth geometries, where the scalar field acts as a
thick Brane.
 We also discuss briefly the graviton localization in our setup and demonstrate the characteristic volcano-like localizing potential for gravitons.
 \end{abstract}
%%%%%%%%%%%%%%%%%%%%%%%%%%%%%%%%%%%%%%%%%%%%%%%%%%%%%%%

\section{Introduction}

 Brane models provide a simple and elegant example of gravitational theories with extra dimensions \cite{RUBSHA} \cite{AHDD} \cite{RS1}. The main concept of
these scenarios is that spacetime has additional space-like dimensions, which may or may not be compact. This higher-dimensional
spacetime, also called the ``bulk", is assumed to contain topological defects of lower dimensionality, where the Standard Model fields are localized.
Although ordinary matter and gauge fields are trapped inside this defect, resembling a membrane situated inside the bulk spacetime,
gravity is free to propagate in the entire extra dimensional surroundings, as well as on the brane itself. The phenomenological implications
of such a setup include modifications of ordinary four-dimensional gravity at small distances and can account for the large disparity between
the weak scale and the Planck mass, providing us with a solution to the hierarchy problem.

One of the most popular brane models is the Randall-Sundrum 2 \cite{RS2} setup, where the spacetime of the theory is five dimensional and there is a single
brane where the Standard Model fields are localized. The extra dimension, denoted by $y$, is not compact and we assume it extends from
$-\infty<y<+\infty$. The brane is located at the origin, $y=0$. Although the extra dimension is infinite in extent, it can be proved that due to
the highly curved geometry of the Randall-Sundrum model, which is essentially an $AdS_5$ space outside the brane, gravity remains highly
localized on the brane and the usual four-dimensional Newton's law is recovered for ordinary distances. For small distances, the presence of the
extra dimension introduces corrections of order $O(\frac{1}{r^3})$.

Our goal is to extend this brane model by including an additional scalar field which is non-minimally coupled to gravity. Although such setups
have already been studied \cite{DLR}, the inclusion of the scalar field is usually treated as a perturbation that has a negligible effect on the geometry. Here
we will solve the complete problem by taking the back-reaction of the scalar field into account and see in what ways this changes the underlying
geometry of spacetime. We will discuss a number of solutions where the geometry is the typical Randall-Sundrum 2 and present the corresponding
profile for the scalar field. We will also demonstrate solutions in the presence of a thick brane induced by the potential of the scalar field itself. In both
cases we will see that there is only a limited range of values for the coupling parameter $\xi$ for which we get physically meaningful results.

%%%%%%%%%%%%%%%%%%%%%%%%%%%%%%%%%%%%%%%%%%%%%%%%%%%%%%%
\section{Mathematical Framework and Equations of Motion}
The action of our model comprises three distinct terms
\begin{equation}
S=S_{gr}+S_{\Phi}+S_{br}
\label{action}
\end{equation}
The first of them accounts for gravity and the non-minimal coupling to the scalar field
\begin{equation}
S_{gr}  = \int {d^5 x} \sqrt { - g} f\left( \Phi  \right)R
\end{equation}
the second is the action of the scalar field
\begin{equation}
S_\Phi   = \int {d^5 x} \sqrt { - g} \left( { - \frac{1}
{2}g^{AB} \nabla _A \Phi \nabla _B \Phi  - V\left( \Phi  \right)} \right)
\end{equation}
and the third is the action of the brane, together with the five-dimensional cosmological constant $\Lambda$
\begin{equation}
S_{br}  = \int {d^5 x} \sqrt { - g} \left( { - \Lambda  - \sigma \delta \left( y \right)} \right)
\end{equation}

Here the brane tension is denoted by $\sigma$, and may be a function of the scalar field $\Phi$, while the
function $f(\Phi)$ measures the strength of the coupling between the scalar field and gravity. It plays the role
of an effective five-dimensional Newton's constant. In order to get back ordinary Einstein gravity we just make
the substitution $f=\frac{1}{16\pi G}$. In the rest of our discussion we will adopt for $f(\Phi)$ the ansatz
$f=1-\frac{1}{2} \xi \Phi^2$. In the units we use, $\xi=0$ corresponds to $f=1$ and ordinary gravity. The
$\xi \Phi^2$ term can account for phenomena such as electroweak symmetry breaking \cite{DLR}.

In order to get the equations of motion for our model we vary the action (\ref{action}). Varying with respect to
the metric $g_{\mu \nu}$ yields Einstein's equations

\begin{equation}
G_{\mu \nu }  = 8\pi GT_{\mu \nu }
\end{equation}
while varying with respect to the field $\Phi$ we get the following equation for the scalar

\begin{equation}
g^{AB} \nabla _A \nabla _B \Phi  - \frac{{\partial V}}
{{\partial \Phi }} + R\frac{{df}}
{{d\Phi }} = 0
\end{equation}
We will use an ansatz for the metric of the form

\begin{equation}
g_{MN}  = \left( {\begin{array}{*{20}c}
   {e^{A\left( y \right)} n_{\mu \nu } } & 0  \\
   0 & 1  \\

 \end{array} } \right)
\end{equation}

and thus we end up with the following three equations of motion

\begin{equation}
f\left( {A'^2  + A''} \right) =  - A'f' - \frac{2}
{3}f'' - \frac{1}
{6}\Phi '^2  - \frac{1}
{3}\left( {\Lambda  + V + \sigma \delta \left( y \right)} \right)
\label{equation-ii}
\end{equation}

\begin{equation}
fA'^2  = \frac{1}
{6}\Phi '^2  - \frac{4}
{3}A'f' - \frac{1}
{3}\left( {\Lambda  + V} \right)
\label{equation-55}
\end{equation}

\begin{equation}
\Phi '' + 2A'\Phi ' - \frac{{\partial V}}
{{\partial \Phi }} - \left( {4A'' + 5A'^2 } \right)\frac{{df}}
{{d\Phi }} = 0
\label{equation-scalar}
\end{equation}

The first two come from Einstein's equations, while the third is the scalar equation. Together they form a
non-linear system of ordinary differential equations. As we can easily prove, only two of the three are
independent, with the third being an expression of energy conservation. Our goal is to solve for the
unknown functions $A$, $\Phi$ and $V$ ($f$ is in our case a known function of $\Phi$ of the form $1-\frac{1}{2} \xi \Phi^2$).
Here $A$ is the function which determines the geometry, $\Phi$ is the scalar field and $V$ the scalar potential.
Since we only have two independent differential equations and three unknown functions, our system is underdetermined.
We thus have the freedom to choose a fitting solution for one of the functions and then solve for the other two.
The system of equations is also complemented by the following junction conditions for the derivatives of $A$ and $\Phi$ on
the brane

\begin{equation}
\left. {A'} \right|_{ - \varepsilon }^\varepsilon   =  - \frac{\sigma }
{{3f\left( 0 \right) + 8\dot f^2 \left( 0 \right)}}
\label{junction-1}
\end{equation}

\begin{equation}
\left. {\Phi '} \right|_{ - \varepsilon }^\varepsilon   =  - \frac{{4\sigma \dot f\left( 0 \right)}}
{{3f\left( 0 \right) + 8\dot f^2 \left( 0 \right)}}
\label{junction-2}
\end{equation}
where we used the abbreviation $\dot f\left( 0 \right) = \left. {\frac{{df}}{{d\Phi }}} \right|_{y = 0} $.

%%%%%%%%%%%%%%%%%%%%%%%%%%%%%%%%%%%%%%%%%%%%%%%%%%%%%%%

\section{Exact Solutions for a RS-2 geometry}
As our equations are coupled and non-linear, solving the system analytically is not a trivial task. One idea is to
use numerical evaluation for some particular form of the scalar potential $V$ \cite{FP}. We will instead keep the potential
arbitrary and try to solve analytically for specific choices for one of the remaining functions $A$ and $\Phi$.
We will then derive the resulting potential and check whether the results we get are physically acceptable.
Combining equations (\ref{equation-ii}) and (\ref{equation-55}) we can eliminate $V$ and thus get the single
equation

\begin{equation}
 - \frac{4}
{3}A'f' + fA'' =  - A'f' - \frac{2}
{3}f'' - \frac{1}
{3}\Phi '^2  - \frac{1}
{3}\sigma \delta \left( y \right)
\label{master}
\end{equation}

We now go on to choose $A(y)$ and solve for $\Phi(y)$ from equation (\ref{master}). Our first choice is the
Randall-Sundrum-2 geometry function $A=-2ky$ with $f = 1 - \frac{1}{2}\xi \Phi ^2 $, so we get a second
order ordinary differential equation involving only $\Phi(y)$

\begin{equation}
 - 2\xi \frac{{\Phi ''}}
{{\Phi '}} + \left( {1 - 2\xi } \right)\frac{{\Phi '}}
{\Phi } - 2\xi k = 0
\label{ode1}
\end{equation}
The solution to this equation is of the form

\begin{equation}
\Phi \left( y \right) = \Phi (0) \left( {1 +C \left( {1 - e^{ - \frac{ky}{2}} } \right)} \right)^{\frac{{2\xi }}
{{4\xi  - 1}}}
\label{solution1}
\end{equation}

 \begin{figure}[!h]
\centering
   \begin{minipage}[c]{0.5\textwidth}
  \centering  \includegraphics[width=1\textwidth]{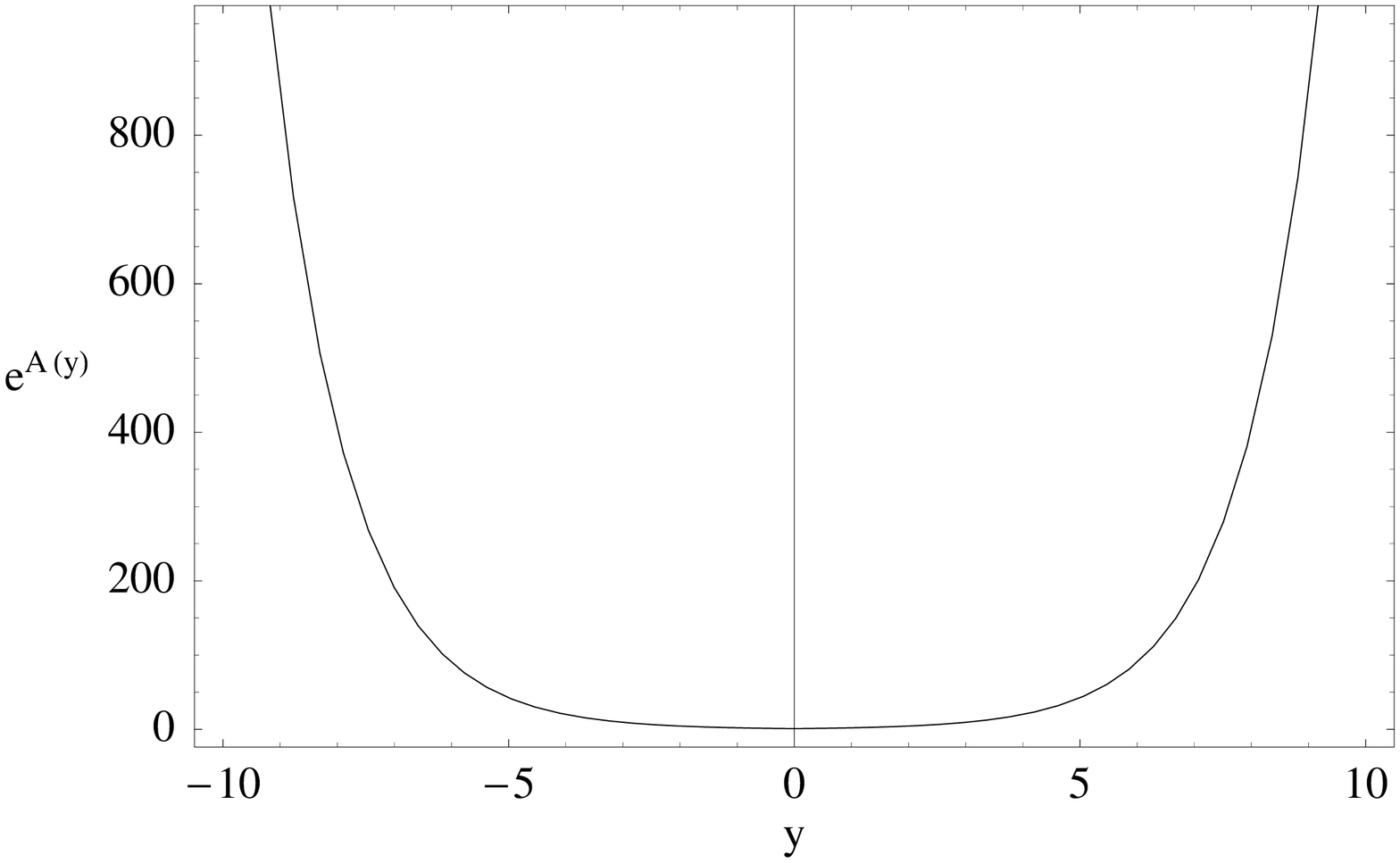}
  \caption{Warp factor for $\xi = \frac{3}{16}$. }
\label{figure1}
   \end{minipage}%
   \begin{minipage}[c]{0.5\textwidth}
  \centering  \includegraphics[width=1\textwidth]{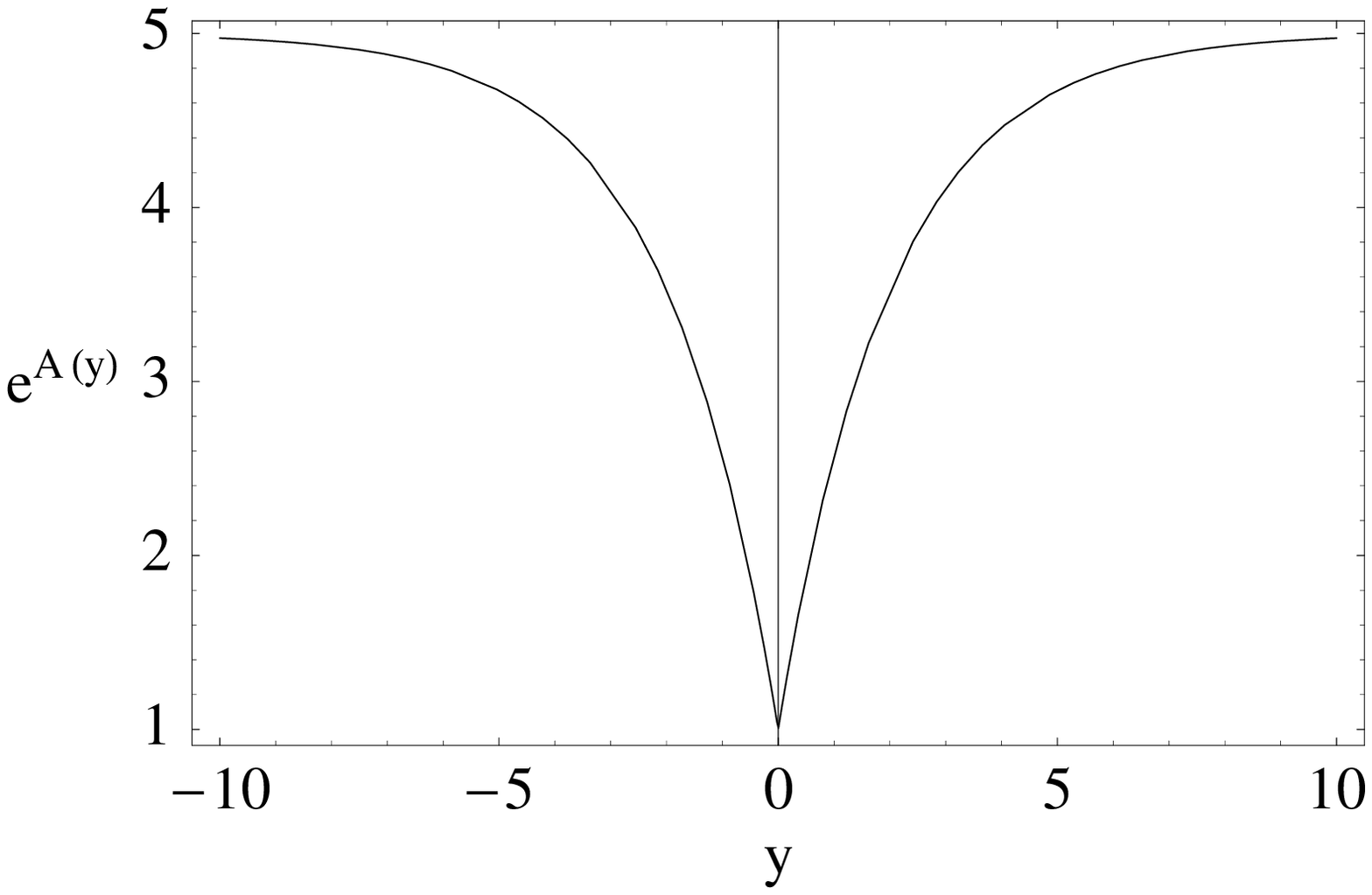}
  \caption{Warp factor for $\xi >\frac{3}{16}$.}
\label{figure2}
   \end{minipage}%

   \label{figure1-2}
\end{figure}%

Plots for this scalar field with respect to $y$ are presented in figures (\ref{figure1}) and (\ref{figure2}). We see that for values of the coupling
parameter $\xi>\frac{3}{16}$, the scalar field reaches asymptotically a constant value and has a profile similar to a folded kink. At the conformal value $\xi=\frac{3}{16}$, the field becomes exponentially increasing, blowing up
at infinity while for $\xi<\frac{3}{16}$, there is a point before infinity where $\Phi$ assumes an infinite value.
Apparently only solutions within the range $\xi > \frac{3}{16}$ are well-behaved in a single-brane
setup. Solutions with negative $\xi$ also exist. However, deriving the potential corresponding to these, we obtain negative powers of $\Phi$
for $\xi<0$. Therefore solutions with negative $\xi$, although they seem initially acceptable, are excluded
because they lead to singular potentials. For the conformal value of the coupling parameter we get a constant
potential.

Until now we have assumed a constant brane tension $\sigma$. For a $\sigma(\Phi)$ brane tension the resulting junction conditions are

\begin{equation}
\left. {A'} \right|^\varepsilon  _{ - \varepsilon }  =  - \frac{{\sigma  + 2\dot f\left( 0 \right)\dot \sigma _0 }}{{3f\left( 0 \right) + 8\dot f^2 \left( 0 \right)}}
\end{equation}

\begin{equation}
\left. {\Phi '} \right|^\varepsilon  _{ - \varepsilon }  =  - \frac{{4\sigma \dot f\left( 0 \right) - 3f\left( 0 \right)\dot \sigma _0 }}{{3f\left( 0 \right) + 8\dot f^2 \left( 0 \right)}}
\end{equation}
and the corresponding solution is of the form

\begin{equation}
\Phi \left( y \right) = \Phi \left( 0 \right)\left( {1 + \frac{{\left( {4\xi  - 1} \right)\left( {\dot \sigma _0  + 16k\xi \Phi _0 } \right)}}{{4k\xi \Phi _0 }}\left( {1 - e^{ - \frac{k\left| y \right|}{2}} } \right)} \right)^{\frac{{2\xi }}{{4\xi  - 1}}}
\end{equation}
We see that we get a similar profile as before. The novel feature is the presence of the derivative ${\dot \sigma _0 }$ of the brane tension with respect to $\Phi$ at the origin. Because of this extra term, an exponentially
decreasing behavior is also possible in this setup.

\section{Solutions for a Smooth Space}
\begin{figure}[!h]
\centering
   \begin{minipage}[c]{0.5\textwidth}
  \centering  \includegraphics[width=1\textwidth]{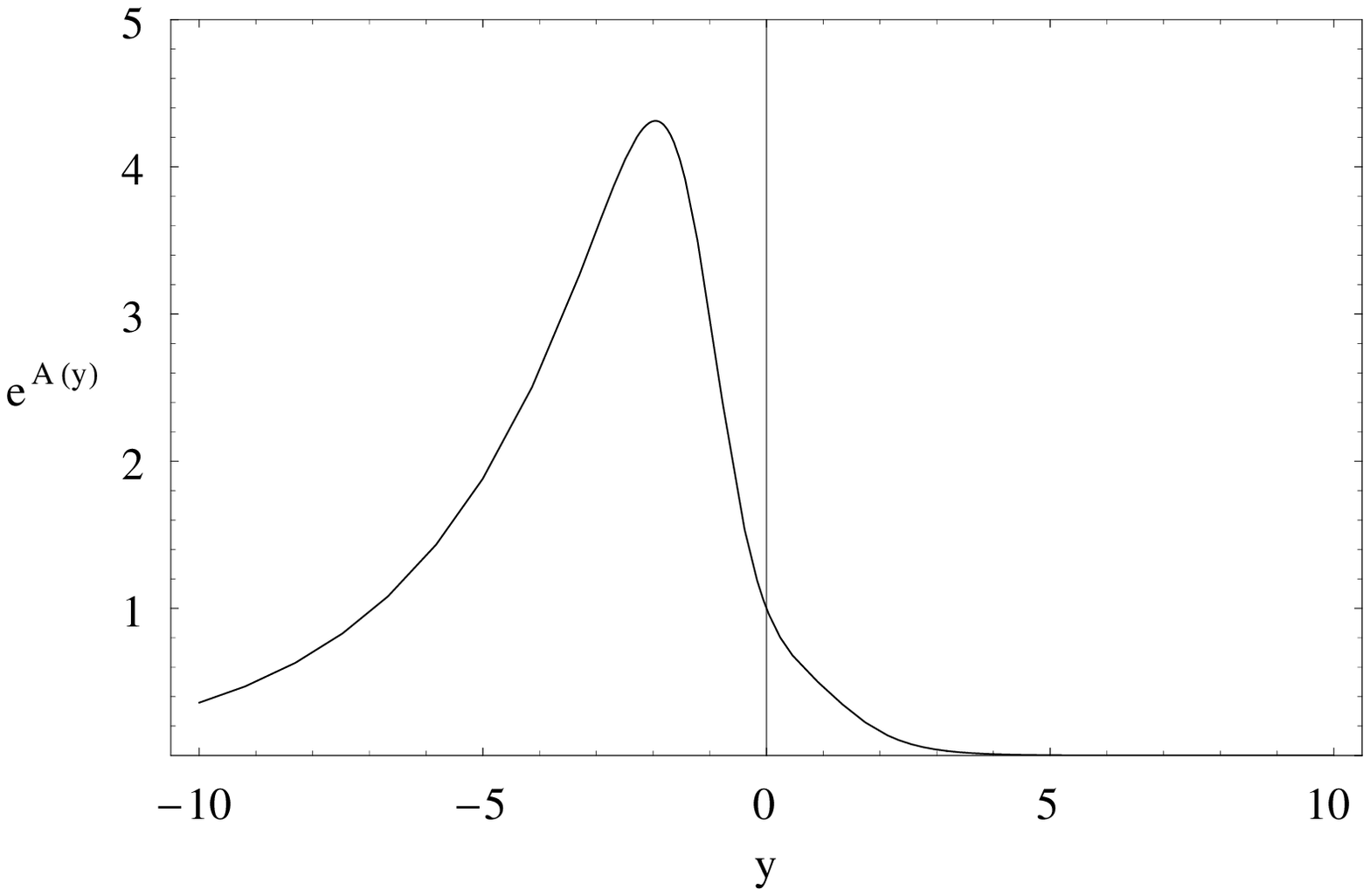}
  \caption{Warp factor for $\xi = 1.6$.}
   \label{figure3}
   \end{minipage}%
   \begin{minipage}[c]{0.5\textwidth}
  \centering  \includegraphics[width=1\textwidth]{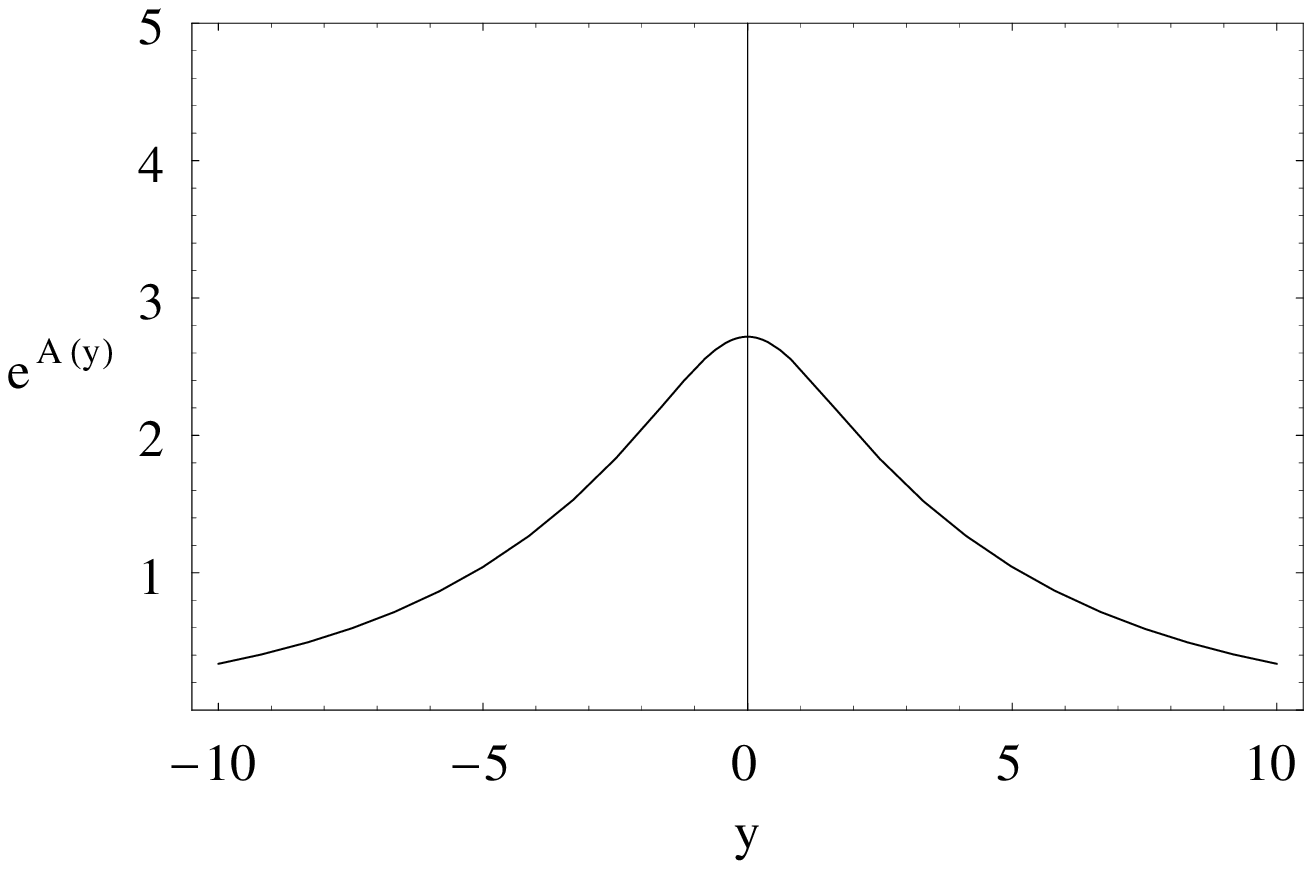}
  \caption{Warp factor for $\xi = 0.2$.}
   \label{figure4}
   \end{minipage}%

\end{figure}

In the previous section we chose a geometry and derived the scalar field and its potential. Here we will work
the other way around, choosing a specific profile for $\Phi$ and finding a suitable warp factor. Since we just
saw that the solutions for the scalar exhibit a folded kink-like profile, we will try to see if an exact kink is actually
a solution. We thus set

\begin{equation}
\Phi \left( y \right) = \frac{{\Phi '_0 }}{a}\tanh \left( {ay} \right)
\end{equation}
and solve the corresponding differential equation. First we will find some numerical solutions for different values of the coupling parameter. Now we do not
have a brane and we do not impose $Z_2$ symmetry. We adopt two different sets of initial conditions, with $A(0)=0$, $A'\left( 0 \right) =  - 1$, which is similar to the Randall-Sundrum case and with $A(0)=1$, $A'\left( 0 \right) =  0$ \cite{KT}. After numerically solving with Mathematica, the
former set yields the solutions of the form seen in figure (\ref{figure3}), while the later gives the results seen in figure  (\ref{figure4}). In the
first case the warp factor is not symmetric about $y=0$ and it remains finite only for a range of values of $\xi$. In the second case, the warp factor is symmetric and finite for all $\xi$. It is evident from the profile of
the warp factor that this geometry resembles a smooth brane, created from the presence of the scalar field.

We can also obtain analytical solutions if we fine-tune our parameters. We can see that

\begin{equation}
e^{A\left( y \right)}  = \left( {\cosh \left( {ay} \right)} \right)^{ - \gamma },\,\,\,\Phi \left( y \right) = \Phi _0 \tanh \left( {ay} \right)
\end{equation}
is a solution provided that the parameters satisfy the following conditions

\begin{equation}
\gamma  = 2\left( {\xi ^{ - 1}  - 6} \right),\,\,\,\Phi _0  = a^{ - 1} \dot \Phi \left( 0 \right) = \left( {2M^3 } \right)^{1/2} \sqrt {\frac{{6\left( {1 - 6\xi } \right)}}{{\xi \left( {1 - 2\xi } \right)}}},\,\,\,
0< \xi < 1/6
\end{equation}
 \begin{figure}[!h]
\centering
   \begin{minipage}[c]{0.5\textwidth}
  \centering  \includegraphics[width=1\textwidth]{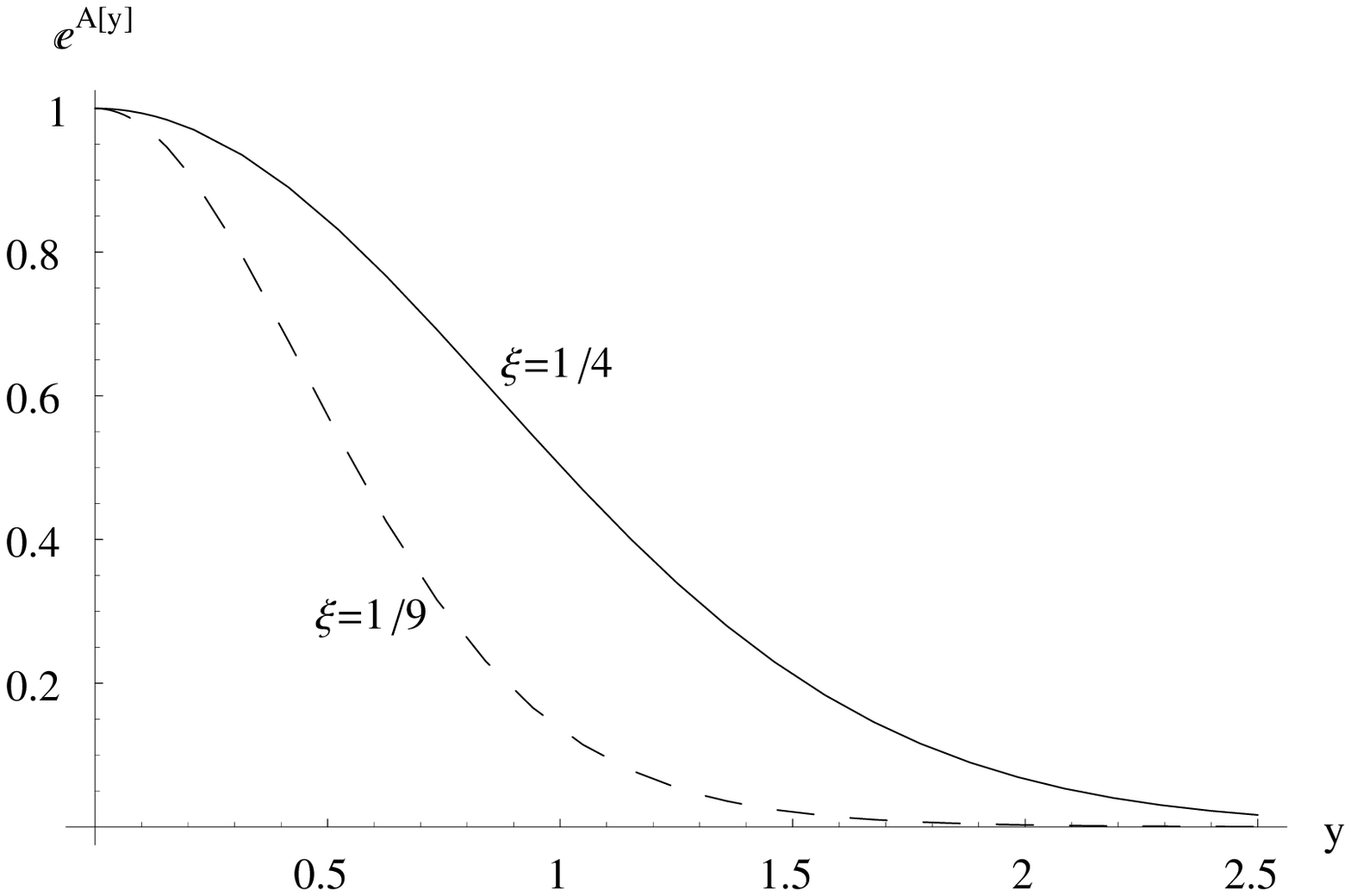}

   \end{minipage}%
   \begin{minipage}[c]{0.5\textwidth}
  \centering  \includegraphics[width=1\textwidth]{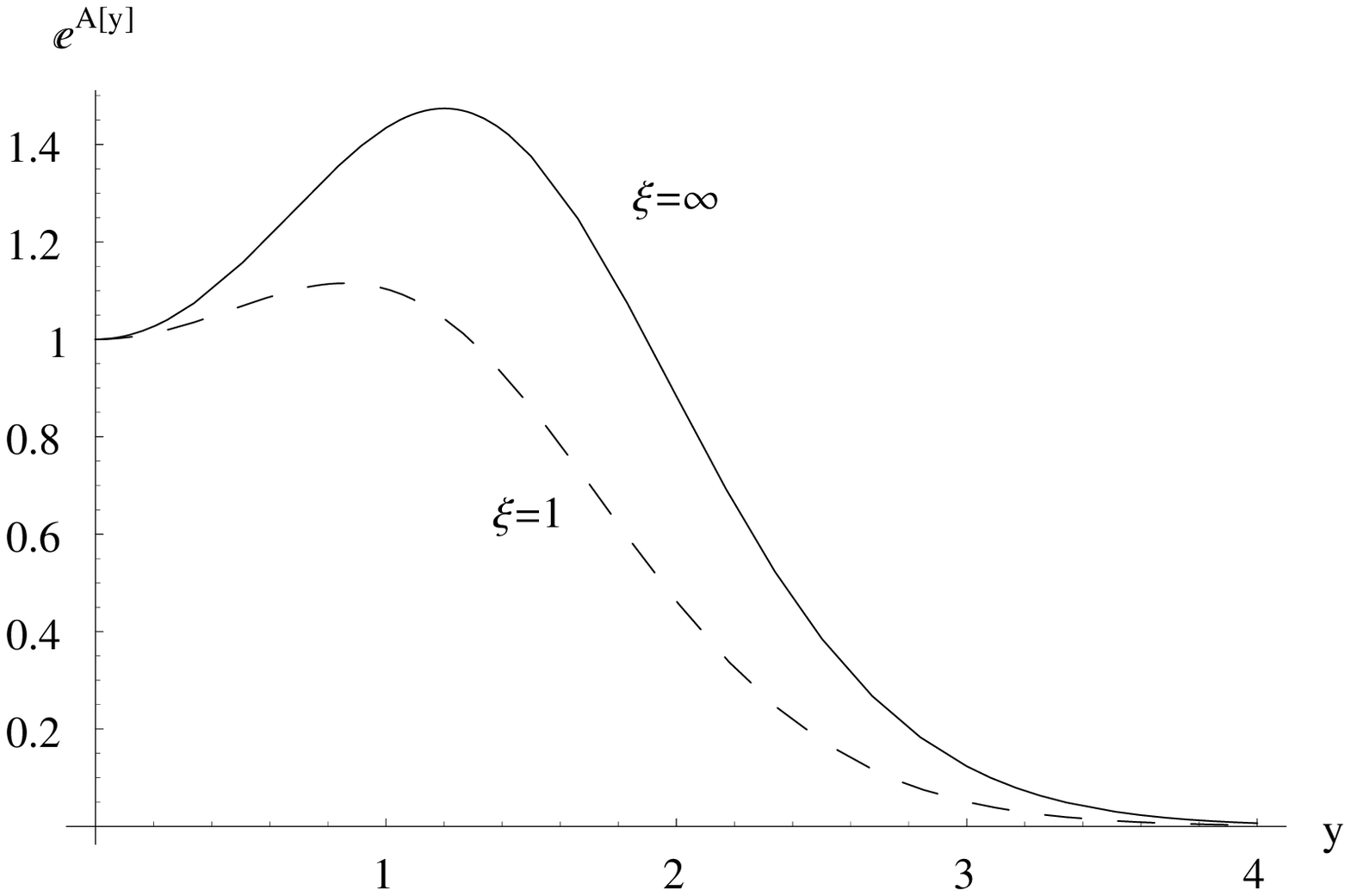}

   \end{minipage}%

     \caption{Warp factors for a kink-like scalar with fine-tuned parameters.}

   \label{figure5}
\end{figure}%
For a different fine-tuning we obtain the solution

\begin{equation}
{\rm A}\left( y \right) =  - 4\ln \left( {\cosh \left( {ay} \right)} \right) + \frac{1}{3}\left( {8 - \xi ^{ - 1} } \right)J\left( {\tanh ^2 \left( {ay} \right)} \right),\,\,\,\Phi \left( y \right) = \Phi _0 \tanh \left( {ay} \right)
,\,\,\, \Phi _0  = 2\sqrt {\xi ^{ - 1} M^3 }
\end{equation}
with
\begin{equation}
J\left( x \right) = \int {dx\left( {1 - x} \right)^{ - 2/3} {}_2F_1 \left( {1/2,1/3,3/2,x} \right) = xF_{PFQ} \left( {\{ 1,1,7/6\} ,\{ 3/2,2\} ,x} \right)}
\end{equation}
Profiles of the warp factor are shown in figure (\ref{figure5}). We see that for $\xi>\frac{1}{2}$ a peak develops away
from the origin.

\section{Graviton Localization}
In this final section we will deal with the issue of graviton localization in the presence of the non-minimally coupled
scalar. Gravitons are described by the perturbations of the metric tensor, so we perturb the metric and leave the
scalar field unchanged

\begin{equation}
\delta g_{{\rm M}{\rm N}}  = \delta ^\mu  _M \delta ^\nu  _N h_{\mu \nu } \left( {x,y} \right),
 \,\,\,\,\,\,\,\,\,\,   \delta \Phi  = 0\,,
\end{equation}
Working in the transverse-traceless gauge  $h^\mu  _\mu   = \partial _\mu  h^{\mu \nu }  = 0$ we derive the equation of motion for the graviton

\begin{equation}
\left( { - \frac{{d^2 }}
{{dy^2 }} + \ddot A(y) + {\dot{A}}^2(y)} \right)\psi(y)  =
 m^2 e^{ - A(y)} \psi(y)\,
\end{equation}
With a proper change of variables this can be put into the form of a one-dimensional Schrodinger equation

\begin{equation}
\left(-\frac{d^2}{dz^2}+U(z)\,\right)\overline{\psi}=m^2\overline{\psi}\,
\end{equation}
with the localizing potential
\begin{equation}
U(z)=\frac{3}{4}\frac{d^2A}{dz^2}+\frac{9}{16}\left(\frac{dA}{dz}\right)^2\,
\end{equation}

 \begin{figure}[!h]
\centering
   \begin{minipage}[c]{0.5\textwidth}
  \centering  \includegraphics[width=1\textwidth]{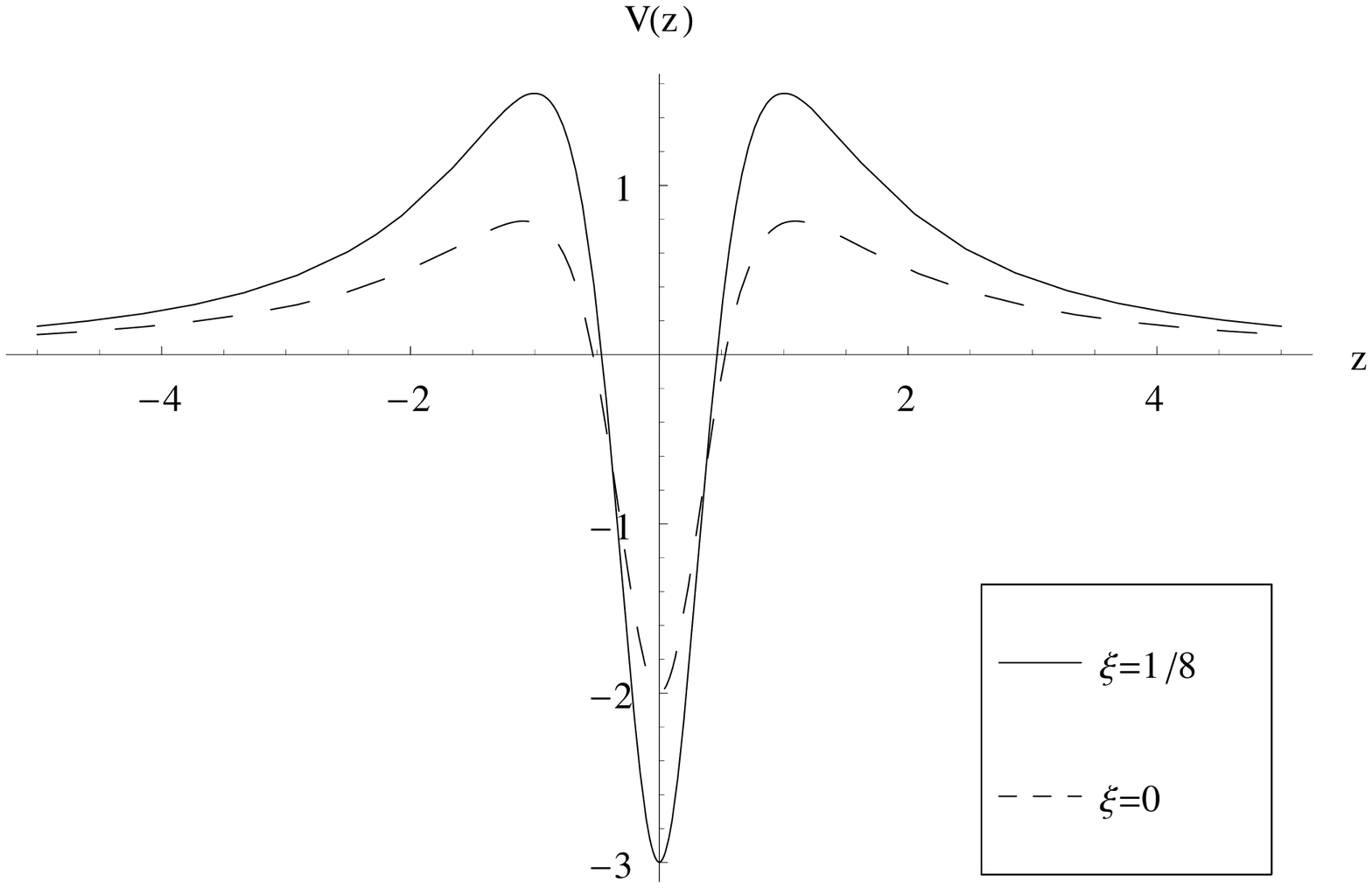}

   \end{minipage}%
   \begin{minipage}[c]{0.5\textwidth}
  \centering  \includegraphics[width=1\textwidth]{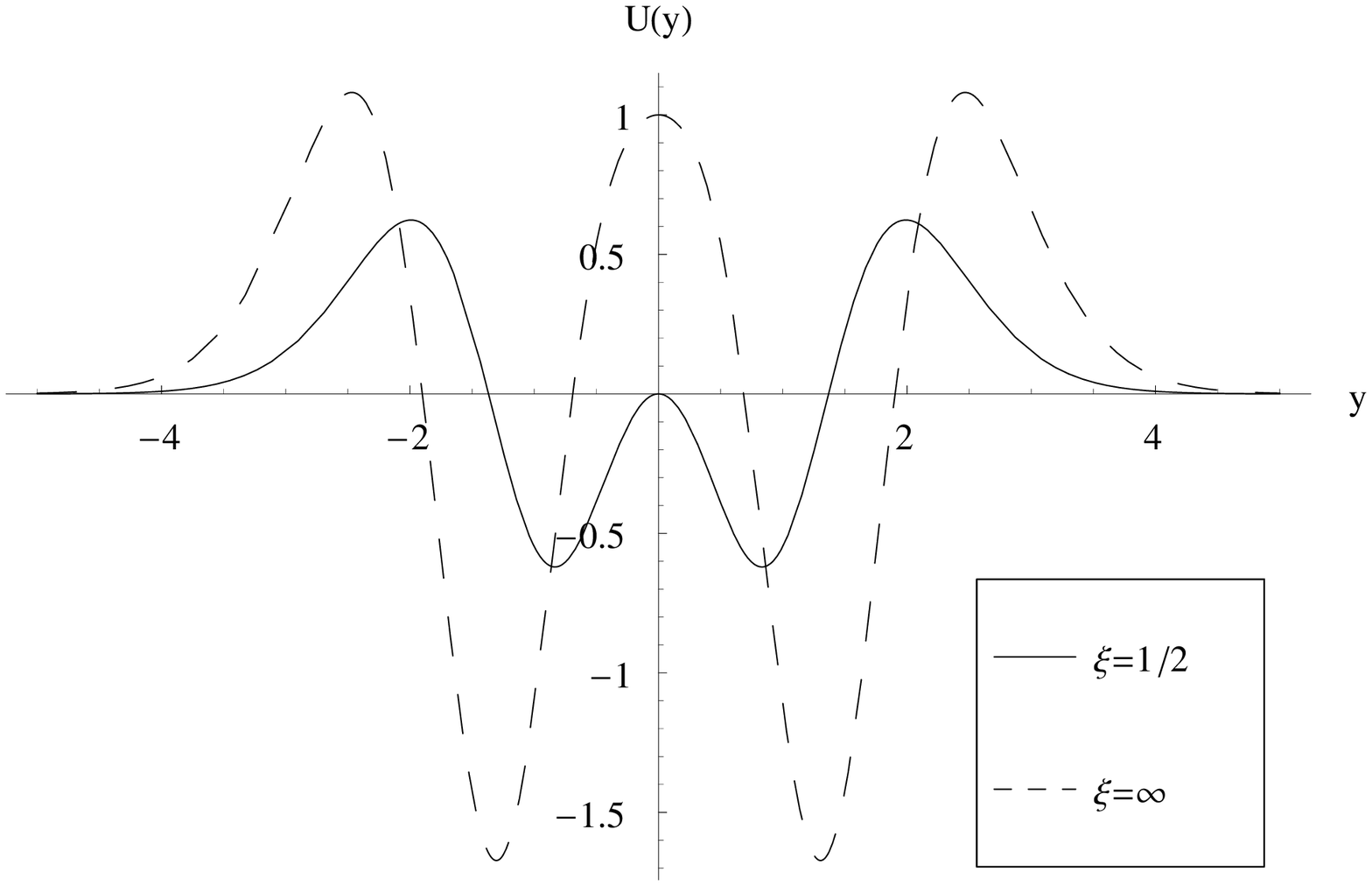}

   \end{minipage}%

   \caption{Localizing potentials for different values of $\xi$. }
   \label{figure6}
\end{figure}%

Because the operator acting on the graviton wavefunction is positive-definite, no tachyon modes are present.
There is also a normalizable zero mode which reproduces ordinary four-dimensional gravity on the brane.
Since for large distances the potential behaves as $e^A$ and this must reduce to zero for reasonable
geometries, we conclude that there is no mass gap in the spectrum of modes and the continuum starts at
$m=0$. The form of the potential is shown in figure (\ref{figure6}) for the warp factors we derived in the previous section. We see that it has the familiar volcano-like profile known from the Randall-Sundrum model. However, this profile
changes and a peak at $y=0$ develops as $\xi$ increases. For $\xi=1/2$ the potential at the origin becomes equal
to zero and it approaches unity as $\xi \to \infty$. We find that the peak first occurs for $\xi=\frac{1}{16}(\sqrt{193}-9)$.

\section{Conclusions}
We presented analytical and numerical solutions for brane models with a non-minimally coupled scalar field. We
found solutions for the scalar field resembling a folded kink, which are compatible with the Randall-Sundrum model. For the case of an exact kink scalar field we derived the corresponding geometries and showed that they are similar
to smooth brane spaces and can sustain graviton localization for the zero mode near the origin.

\ack This presentation is based on work done in collaboration with A. Dimitriadis and K. Tamvakis at the University
of Ioannina, co-funded by the European Union
in the framework of the Program $\Pi Y\Theta A\Gamma O PA\Sigma-II$
of the {\textit{``Operational Program for Education and Initial
Vocational Training"}} ($E\Pi EAEK$) of the 3rd Community Support Framework
of the Hellenic Ministry of Education, funded by $25\%$ from
national sources and by $75\%$ from the European Social Fund (ESF). C. B.
aknowledges also an {\textit{Onassis Foundation}} fellowship.

\end{document}